\documentclass[a4paper,english]{article}
\usepackage[latin2]{inputenc}
\usepackage[T1]{fontenc}
\usepackage[intlimits]{amsmath}
\usepackage{amsfonts}
\usepackage{times}
\usepackage{graphicx}
\usepackage{makeidx}
\usepackage{amssymb}

\newcommand{\nn}{\nonumber}

\begin{document}
\large\rm
 \renewcommand\figurename{}

\renewcommand\tablename{\large\sf Tab.}

\title{\LARGE\textbf{Higgs Scalar-Tensor Theory for Gravity and the Flat
Rotation Curves of Spiral Galaxies}}
\author{Nils M. Bezares Roder\footnote{Institut für Theoretische Physik,
Universität Ulm},
Heinz Dehnen\footnote{Fachbereich Physik, Universität Konstanz}}

\date{Ulm, \today}



\thispagestyle{empty} \vspace*{2.7cm}
\begin{center}\textbf{\Large{Higgs Scalar-Tensor Theory for Gravity and the
Flat Rotation Curves of Spiral Galaxies}}\\
\vspace{0.5cm}\textsc{Nils M. Bezares-Roder$^{1}$, Heinz Dehnen$^{2}$}\\
\emph{1. Institut für Theoretische Physik, Universität
Ulm}\footnote{Albert-Einstein-Allee 11, D-89069 Ulm. E-mail: Nils.Bezares@uni-ulm.de}\\
\emph{2. Fachbereich Physik, Universität Konstanz}\footnote{Fach M677
Fachbereich Physik, D-78457 Konstanz. E-mail: Heinz.Dehnen@uni-konstanz.de}\\
\vspace {0.5cm} {\small DOI 10.1007/s10714-007-0449-8\\
General Relativity and Gravitation \textbf{39}(8), 1259-1277
(2007)\footnote{The original publication is available at
www.springer.com.}}
\end{center}

\date{\today}
\begin{center}\textbf{Abstract}\\
The scalar-tensor theory of gravity with the Higgs field as scalar
field is presented. For central symmetry it reproduces the
empirically measured flat rotation curves of galaxies. We
approximate the galaxy by a polytropic gas sphere with the
polytropic index $\gamma=2$ and a massive core.\\\end{center}

\section{Introduction}
Usually, flat rotation curves of spiral galaxies are interpreted in
such a way that the galaxies contain a significant quantity of
non-luminous (dark) matter (DM), which leads to a great mass
discrepancy $D=\frac{M_{dyn}}{M_{obs}}-1$ between the luminous
-observed- and dynamical mass of a galaxy. Another interpretation is
that the usual Newtonian inverse square law of gravity has to
be modified on large scales.\\
Currently, the nature and the distribution of dark matter is
unknown. It is suggested that a spiral galaxy is surrounded by a
spherical halo of cold dark matter (CDM), the extension of which is
much greater than the visible disk \cite{Ostriker73}. This idea is
determined from measurements of the 21-cm line of neutral hydrogen
outside the region of the optical radius. Additionally, these
measurements point out that there exists a sensitive coupling
between visible and dark matter \cite{Albada88} (disc-halo
conspiracy). Today, the dark matter hypothesis is generally
accepted, mainly due to the fact that changing established physical
laws is usually avoided.\\
Nevertheless, the Newtonian $r^{-2}$-law is only empirically tested
in the satellite region and the solar system, so that flat rotation
curves might be a first hint for its failing on large scales ($\geq
20 kpc$). In this context, several modifications of the Newtonian
law have been proposed, but nearly all of them are constructed
\emph{ad hoc}.\\
One important attempt of modification is the FLAG (\textbf{f}inite
\textbf{l}ength-scale \textbf{a}nti-\textbf{g}ravity)-theory of
Sanders \cite{Sanders86}, who added a Yukawa potential with finite
length scale $r_0$ and a negative coupling constant $\hat{\alpha}$
to the usual gravitational potential
\begin{align}
\Phi= -\frac{G_\infty M}{r}(1+ \hat{\alpha} e^{-r/r_0}),\label{FLAG}
\end{align}
where $G_\infty$ is the value of the gravitational constant at
infinity, whereas the local value is variable. For $\hat{\alpha}
\approx -0,92$ and 20kpc $<r_0< 40$kpc, the model reproduces the
rotation curves of galaxies ranging in size from 5 to 40kpc.\\
Milgrom \cite{Milgrom83} found that the deviation from the Newtonian
law in large astronomical systems apparently appears below a
critical acceleration. In his \textbf{mo}dified \textbf{N}ewton
\textbf{d}ynamics (MOND) the equation of motion becomes
\begin{align}
m \mu(a/a_0)\vec{a}= \vec{F}, \label{MOND}
\end{align}
where $a_0$ is a critical acceleration and $\mu(a/a_0)$ is a
function determined from observations and having the asymptotical
behaviour $\mu(a/a_0)=a/a_0$ for $a\leq a_0$ and $\mu(a/a_0)=1$ for
$a\gtrsim a_0$. Equation (\ref{MOND}) leads to acceptable results
but does not have a deeper theoretical foundation. Therefore,
Beckenstein and Milgrom \cite{Beckenstein84} tried to formulate the
MOND within a field theory. They started from a non-relativistic
Lagrangian, which leads to a modified Poisson equation. In this way
they can describe mass discrepancies in low accelerated systems.\\
Additionally, there exist other field theories of gravity that
modify even the classical General Relativity by the addition of one
or more scalar fields to the tensor field of General Relativity.
This kind of general relativistic models with a scalar field are
conform equivalent with more-dimensional general relativistic models
\cite{Cotsakis97}. The first such attempt was started by Jordan
\cite{Jordan55}. He noticed through his isomorphy theorem that
projective spaces as Kaluza-Klein's can be reduced to usual
Riemannian 4-dim spaces and that a scalar field as fifth component
of such a projective metric can play the role of a variable
gravitational ``constant'', by which it is possible to vary
the strength of gravitation \cite{Fauser01}.\\
In his theory, Jordan introduced two coupling parameters of the
scalar field, one producing a variation of the gravitational
constant and another one that would break the energy conservation
through a non-vanishing divergence of the energy-stress tensor to
increase the mass in time, in accordance with the ideas of Dirac
\cite{Jordan68}. However, the microwave background radiation as a
real black-body radiation discovered in 1965 forces us to accept
general energy conservation as given \cite{Hoenl68}. Jordan's theory
was worked out independently by Brans and Dicke \cite{Brans61}
without changing energy conservation, but again introducing a scalar
field with an infinite length scale and playing the role of a
variable gravitational coupling. This time it followed Mach's
principle of the relativity of inertia, that with respect to the
equivalence principle, the inertial, as well as passive and active
gravitational mass should be induced by the interaction with the
gravitational field \cite{Einstein13}. O. Obregón investigated in
that subject beginning in the 1970's. The only coupling parameter
$\omega$ of this scalar-tensor theory is a measure of the strength
of the scalar field coupling to matter. For a sensitive theory, it
should be of the order of unity \cite{Brans61}. For $\omega
\rightarrow \infty$, General Relativity is obtained, and, in fact,
solar system experiments restrict $\omega$ to be greater than about
500 \cite{Reasensberg79}, which entails that the Brans-Dicke theory
leads nearly to the same results as General Relativity.\\
On the other hand, in the elementary particle physics the inertial
as well as passive gravitational masses are generated simultaneously
with respect to gauge invariance by the interaction with a scalar
Higgs field through the so-called Spontaneous Symmetry Breakdown.
Because of the identity of passive and active gravitational mass
also the latter, i.e. the gravitational constant should be produced
by the same symmetry breaking. Additionally, Dehnen et al.
\cite{Dehnen90, Dehnen91} showed that the interaction of the Higgs
field with the particles which gain mass through it is of a
gravitational and Yukawa nature, so that they finally identified the
Higgs field with that of the scalar-tensor theory of gravity
possessing a variable gravitational ``constant'' \cite{Dehnen92}. In
such a theory, the Newtonian gravitational constant as well as the
elementary particle masses can be generated simultaneously by
breaking symmetry. Here, the scalar field has a finite length scale,
which causes a Yukawa-like potential similar to the one in
Eq.(\ref{FLAG}), and following the ideas of Zee \cite{Zee79}, as a
broken-symmetric theory of gravity.\\\\

Gessner showed \cite{Gessner92} that a negative cosmological
constant $\Lambda$ gives rise to flat rotation curves that would,
however, reduce the age for the universe too much. Nevertheless, the
scalar-tensor theory of Dehnen \emph{et al.} contains a cosmological
function $\Lambda(\varphi)$ (instead of a constant), proportional to
the Higgs potential $V(\varphi$), and therefore, the above-mentioned
problem may not appear using this theory. Also the nature and value
of the cosmological ``constant'' may be understandable through it.
The first such attempt within the HSTT was made by Cervantes-Cota
and Dehnen \cite{Cervantes95, Cervantes95a} for an explanation of
inflation.

\section{Higgs scalar-tensor theory of gravity}

The model starts with a Lagrange density in unique form ($\hbar=1$,
$c=1$)\footnote{$_{;\mu}$ means the covariant derivative with
respect to all gauged groups.}:
\begin{align}
{\cal L}= \left[\frac{1}{16\pi}\breve{\alpha} \phi^\dagger \phi R+
\frac{1}{2} \phi^\dagger_{;\mu} \phi^{;\mu}-
V(\phi)\right]\sqrt{-g}+ {\cal L}_M\sqrt{-g},\label{Lagrangian}
\end{align}
where $\phi$ is an $U(N)$ iso-vector, $R$ the Ricci scalar,
$\breve{\alpha}$ a dimensionless factor and ${\cal L}_M$ the
Lagrange density of the fermionic and massless bosonic fields:
\begin{align}
{\cal L}_M= \frac{i}{2}\bar{\psi}\gamma^\mu_{L,R}\psi_{;\mu}+ h.c.-
\frac{1}{16\pi} F^a_{\mu \nu} F^{\mu \nu}_a- k\bar{\psi}_R
\phi^\dagger \hat{x}\psi_L + h.c.\label{Lag-M}
\end{align}
with the Yukawa coupling operator $\hat{x}$. Furthermore, there is a
Higgs potential ($\mu^2<0$, $\lambda >0$ real valued
constants)\footnote{The potential $V(\phi)$ is normalized to zero
for the ground-state. Otherwise, a formal cosmological constant
appears, whereas for Eq.(\ref{Higgspot}) only a cosmological
function exists.}
\begin{align}
V(\phi)= \frac{\mu^2}{2}\phi^\dagger \phi+
\frac{\lambda}{24}(\phi^\dagger\phi)^2+
\frac{3}{2}\frac{\mu^4}{\lambda}.\label{Higgspot}
\end{align}
The field equations for gravity and the Higgs field following from
Eq.(\ref{Lagrangian}) are
\begin{alignat}{1}
R_{\mu \nu}&- \frac{1}{2}Rg_{\mu \nu}+ \frac{8\pi}{\breve{\alpha}
\phi^\dagger \phi} V(\phi)g_{\mu \nu}= -\frac{8\pi}{\breve{\alpha} \phi^\dagger \phi}T_{\mu \nu}-
\frac{4\pi}{\breve{\alpha} \phi^\dagger
\phi}\left[\phi^\dagger_{;\mu}\phi_{;\nu}+
\phi^\dagger_{;\nu}\phi_{;\mu}\right]+\label{Einstein}\\
&+ \frac{4\pi}{\breve{\alpha} \phi^\dagger \phi}
\phi^\dagger_{;\lambda} \phi^{;\lambda} g_{\mu \nu}-
\frac{1}{\phi^\dagger \phi}\left[(\phi^\dagger \phi)_{;\mu ;\nu}-
(\phi^\dagger
\phi)^{;\beta}\,_{;\beta} g_{\mu \nu}\right],\nn\\
\phi^{;\mu}\,_{;\mu}&- \frac{1}{8\pi} \breve{\alpha} \phi R+
\mu^2\phi+ \frac{\lambda}{6} (\phi^\dagger \phi)\phi
=-2k\bar{\psi}_R\hat{x}\psi_L,\label{Higgs}
\end{alignat}
where $T_{\mu \nu}$ is the energy-momentum tensor belonging to the
matter Lagrangian ${\cal L}_M$ in Eq.(\ref{Lagrangian}). It
satisfies the conservation law:
\begin{align}
T_\mu\,^\nu\,_{;\nu}=k\bar{\psi}_R \phi^\dagger_{;\mu} \hat{x}
\psi_L+h.c.+F^a_{\mu \nu} j^\nu_a(\phi).\label{conservation}
\end{align}
$j_a^\nu(\phi)$ are the currents of the scalar field $\phi$. The
coupling operator $\hat{x}$ is not zero in the case of a coupling of
the scalar field with the fermionic field in the matter Lagrangian
${\cal L}_M$ in Eq.(\ref{Lag-M}). Otherwise, it is zero.\\
Instead of $\phi$ we go on to the real-valued function $\varphi$
\begin{align}
\varphi^2=\frac{\phi^\dagger \phi}{v^2},\quad \text{with }
v^2=\phi_0^\dagger \phi_0=-\frac{6\mu^2}{\lambda},
\end{align}
which describes this Higgs field normalized to the ground state
$\phi_0=vN$ ($N^\dagger N=1$) of lowest energy.\\
With the excited field $\xi$ following from
\begin{align}
\varphi^2=1+ \xi
\end{align}
and with
\begin{align}
G=(\breve{\alpha} v^2)^{-1}\label{Gv}
\end{align}
for the gravitational constant, and with the length
scale\footnote{In view of the structure of $l$ with the high values
of $\breve{\alpha}$, only large values of the length scale $l$ are
expected (c.f. \cite{Cervantes95, Cervantes95a}). Further, for the
extreme case $l\rightarrow \infty$, the symmetry can stay broken and
the scalar field $\xi$ act antigravitationally in the exact solution
of the statical field equations \cite{Bezares07}, avoiding, for
instance, a Schwarzschild horizon.}
\begin{align}
l^{-2}=16\pi G \frac{\mu^4}{\lambda}\,\left(1+
\frac{4\pi}{3\breve{\alpha}}\right)^{-1}\,=\,-\frac{8\pi}{3}\frac{\mu^2}{\breve{\alpha}}\,\left(1+
\frac{4\pi}{3\breve{\alpha}}\right)^{-1},\quad \quad
\end{align}
the field equations (\ref{Einstein}) and (\ref{Higgs}) with the use
of $\breve{\alpha} \simeq
\left(\frac{M_{Planck}}{M_{Boson}}\right)^2 \gg 1$ (c.f.
Eq.(\ref{Gv})) become
\begin{alignat}{1}
R_{\mu \nu}&- \frac{1}{2}Rg_{\mu \nu}+ \frac{1}{l^2}(1+
\xi)^{-1}\xi\left(1+ \frac{3}{4}\xi\right)g_{\mu \nu}=\nn\\
&= -8\pi G(1+ \xi)^{-1} \left(T_{\mu \nu}- \frac{\hat{
q}}{3}Tg_{\mu \nu}\right)-
(1+ \xi)^{-1}\xi_{,\mu ;\nu},\label{Einsteininf}\\
\xi^{,\mu}\,_{;\mu}&+ \frac{1}{l^2} \xi= \hat{q}\frac{8\pi
G}{3}T.\label{Higgsinf}
\end{alignat}
$\hat{q}$ is an operator with the value 0 for a coupling of
$\phi$ with the fermionic field through ${\cal L}_M$ ($\hat{=}
\hat{x} \neq 0$) and with the value 1 for only coupling through the
curvature scalar $R$ ($\hat{=}  \hat{x}=0$). For the first case, the
source of the scalar field vanishes identically \cite{Dehnen93}.\\

In Eq.(\ref{Einsteininf}) there exists a cosmological function
\begin{align}
\Lambda(\xi)= \frac{8\pi G}{1+ \xi}V(\xi)= 12\pi G
\frac{\mu^4}{\lambda}\frac{\xi^2}{1+\xi},
\end{align}
which vanishes for the ground state $\xi=0$. This is a consequence
of the special choice $V(\phi_0)=0$, according to
Eq.(\ref{Higgspot}) (see footnote 2).\\
In the field equation (\ref{Einsteininf}), it is possible to
eliminate the Ricci scalar $R$ by multiplying Eq.(\ref{Einsteininf})
by $g_{\mu \nu}$. Now one obtains from Eq.(\ref{Einsteininf}) after
the use of (\ref{Higgsinf}):
\begin{alignat}{1}
R_{\mu \nu}- \frac{1}{2l^2}&(1+ \xi)^{-1}\xi\left(1+
\frac{3}{2}\xi\right)g_{\mu
\nu}=\nn\\
&=- 8\pi G(1+ \xi)^{-1}\left[T_{\mu \nu}- \frac{1}{2}\left(1-
\frac{\hat{q}}{3}\right)T g_{\mu \nu}\right]- (1+
\xi)^{-1}\xi_{,\mu ;\nu}.\label{EinsR}
\end{alignat}
For calculating astrophysical problems, the metric for spherical
symmetry
\begin{align}
ds^2= e^\nu (dx^0)^2- e^\lambda dr^2- r^2(d\vartheta^2+
sin^2\vartheta d\varphi^2)
\end{align}
is introduced, and the energy-momentum tensor $T_{\mu \nu}$ is taken
phenomenologically as that of the ideal liquid ($u^\mu$:
4-velocity):
\begin{align}
T_{\mu \nu}= (\varrho+ p)u_\mu u_\nu -pg_{\mu \nu},\quad u^\mu
u_\mu=1,\label{ideal}
\end{align}
with the pressure $p$ and density distribution $\varrho$. Hence, the
field equations yield ($u$ is the radial 3-velocity of the fluid):
\begin{alignat}{1}
&e^{\nu- \lambda}\left(\frac{\nu''}{2}+ \frac{\nu'^2}{4}-
\frac{\nu'\lambda'}{4}+ \frac{\nu'}{r}\right)-
\frac{\ddot{\lambda}}{2}- \frac{\dot{\lambda}^2}{4}+
\frac{\dot{\lambda}\dot{\nu}}{4}+
\frac{1}{2l^2}(1+\xi)^{-1} \xi\left(1+ \frac{3}{2}\xi\right) e^\nu=\nn\\
&\quad = 8\pi G(1+\xi)^{-1} \left[(e^{-\nu}-
u^2e^{-\lambda})^{-1}(\varrho+ u^2pe^{\nu- \lambda})-
\frac{1}{2}\left(1-\frac{\hat{q}}{3}\right)
(\varrho-3p)e^\nu \right] +\nn\\
&\quad + (1 +\xi)^{-1}\left[\ddot{\xi}-
\frac{\dot{\nu}}{2}\dot{\xi}-
\frac{\nu'}{2}e^{\nu- \lambda}\xi'\right],\label{00Einstein}\\
&e^{\lambda- \nu} \left(\frac{\ddot{\lambda}}{2}+
\frac{\dot{\lambda}^2}{4}- \frac{\dot{\lambda}\dot{\nu}}{4}\right)-
\frac{\nu''}{2}- \frac{\nu'^2}{4}+ \frac{\nu'\lambda'}{4}+
\frac{\lambda'}{r}- \frac{1}{2l^2}(1+\xi)^{-1}
\xi\left(1+\frac{3}{2}\xi\right)e^\lambda=\nn\\
&\quad = 8\pi G(1+\xi)^{-1} \left[(e^{-\nu}-u^2
e^{-\lambda})^{-1}{(u^2 \varrho+ pe^{\lambda-
\nu})+\frac{1}{2}\left( 1-\frac{\hat{q}}{3}\right)(\varrho-
3p)e^\lambda}\right]+\nn\\
&\quad + (1+\xi)^{-1} \left[\xi''-
\frac{\dot{\lambda}}{2}e^{\lambda- \nu}\dot{\xi}-
\frac{\lambda'}{2}\xi'\right],\label{11Einstein}
\end{alignat}
\begin{alignat}{1}
& \frac{\dot{\lambda}}{r}= -8\pi G(1+ \xi)^{-1}\left[e^{-\nu}-
u^2e^{-\lambda}\right]^{-1}(p +\varrho)u- (1+ \xi)^{-1}
\left[\dot{\xi}'- \frac{\nu'}{2}\dot{\xi}-
\frac{\dot{\lambda}}{2}\xi'\right],\label{01Einstein}\\
& e^{-\lambda}\left(1+\frac{r}{2}(\nu'-\lambda')\right)-1+
\frac{r^2}{2l^2}(1+ \xi)^{-1} \xi\left(1+ \frac{3}{2}\xi\right)=\nn\\
&\quad = -8\pi G(1+ \xi)^{-1} \left[pr^2+
\frac{1}{2}\left(1-\frac{\hat{q}}{3}\right)(\varrho-
3p)r^2\right]- (1+ \xi)^{-1} re^{-\lambda} \xi',\label{22Einstein}
\end{alignat}
with the Higgs equation for the excited Higgs field:
\begin{align}
\ddot{\xi}e^{-\nu}- \xi''e^{-\lambda}- \frac{\dot{\nu}-
\dot{\lambda}}{2}e^{-\nu}\dot{\xi}- \frac{\nu'-
\lambda'}{2}e^{-\lambda} \xi'- \frac{2}{r}e^{-\lambda}\xi'+
\frac{1}{l^2}\xi =-\hat{q}\frac{8\pi G}{3}(\varrho-
3p).\label{Diracid}
\end{align}
For the following, we take $\hat{q}=1$. The other case can be
found in Dehnen and Frommert \cite{Dehnen93}. There, Higgs particles
only interact through the gravitational channel.

\section{The Linearized Field Equations}

We will examine the field equations for the case $\hat{q}=1$,
i.e., the case of no coupling of $\phi$ through ${\cal L}_M$.
Therefore, the source of this Higgs field does not disappear and the
Higgs particles are able to be generated, in contrast to the case of
a coupling through ${\cal L}_M$ with only gravitationally coupled
Higgs particles \cite{Dehnen93}. Furthermore, for the investigation
of the flat rotation curves of galaxies only the knowledge of the
functions $\nu(r)$ and $\xi(r)$ is necessary. For them we find in
static linear approximation, from
Eqs.(\ref{00Einstein})-(\ref{Diracid}) the field equations for $\nu$
and $\xi$:
\begin{alignat}{1}
\left(\frac{1}{2}\nu''+ \frac{1}{r}\nu'\right)+
\frac{1}{2l^2}\xi =& 8\pi G\left(\frac{2}{3}\varrho+ p\right), \label{EinlinA}\\
\xi''+ \frac{2}{r}\xi'- \frac{1}{l^2}\xi=& -\frac{8\pi
G}{3}(\varrho- 3p).\label{EinlinD}
\end{alignat}

However, in the case of a singular central mass, a non-linearization
of the $\xi$-field is required for the central area because the
effective gravitational coupling parameter $G(\xi)\sim (1+
\xi)^{-1}$ in Eqs.(\ref{00Einstein})-(\ref{22Einstein}) runs to zero
for $r\rightarrow 0$ (see Eq.(\ref{xi-bnu})). For this reason, it is
possible to linearize the field equations in $\nu$ and $\lambda$ but
not in $\xi$ in the centre. This partially linearized equation for
$\nu(r)$ can be written instead of (\ref{EinlinA}) in the form (c.f.
Eq.(\ref{00Einstein}))
\begin{align}
\frac{\nu''}{2}+ \frac{1}{r}\nu'+ \frac{1}{2l^2}\xi
\frac{1+\frac{3}{2}\xi}{1+\xi}= 8\pi G \frac{\frac{2}{3}\varrho+
p}{1+\xi}. \label{TEinA}
\end{align}

In contrast to this, the $\xi$-field equation (\ref{EinlinD}) is
valid without any change (c.f. Eq.(\ref{Diracid})).\\

On the basis of Eqs.(\ref{EinlinA})-(\ref{EinlinD}) we analyze in
the following the rotation curves of galaxies. Of course, besides
these, further investigations within this theory are necessary
concerning, for instance, the gravitational lensing effect and the
Tully-Fisher law for galaxies.

\section{Sphere-like galaxy with a singular mass in the centre}

For simplicity, we choose a mass-sphere (radius $R_0$) with a
central point mass as galaxy model. The pressure of the sphere is
given by a polytropic equation of state and that of the central mass
$\varrho_S=M_S\delta(\vec{x})$ may be given by an extreme equation
of state
\begin{align}
p_S=a\varrho_S= aM_S\delta(\vec{x})\label{extreme}
\end{align}
with $a$ as the pressure parameter ($0<a\leq 1$, but $a\neq
\frac{1}{3}$)\footnote{The case $a=\frac{1}{3}$ must be considered
separately, which is not performed in this paper.}; $M_S$ is the
central mass.\\
We solve the equations (\ref{EinlinA}) up to (\ref{TEinA}) in
several steps:\\
\begin{itemize}
\item The $\xi$-field for $r> R_0$:\\
Outside of the sphere ($r> R_0$), $p=\varrho=0$ is valid. The
general outer solution for the $\xi$-field with the boundary
condition to be zero in infinity ($r\rightarrow \infty$) is given
through
\begin{align}
\xi_a=\frac{A}{r}e^{-r/l}\label{xi-exterior}
\end{align}
with the integration constant $A$.\\\\

\item The potential of the gravitational field for $r>R_0$:\\
Multiplying Eq.(\ref{TEinA}) for the vacuum with $2r^2$ gives
\begin{align}
(r^2 \nu')'+ \frac{r^2}{l^2}\xi \frac{1+ \frac{3}{2}\xi}{1+ \xi}=
0.\label{r2n'0}
\end{align}
The quotient $\left(1+\frac{3}{2}\xi\right)/(1+ \xi)$ has the value
$\frac{3}{2}$ for $r\rightarrow 0$ and 1 for $r\rightarrow \infty$
and is everywhere monotonous. For this reason, it may be possible to
approximate it by one. Therefore, it follows according to
Eq.(\ref{xi-exterior}):
\begin{align}
(r^2\nu_a')'+ \frac{1}{l^2}Are^{-r/l}=0,
\end{align}
After integration one finds
\begin{align}
\nu'_a=\frac{A}{r^2}e^{-r/l}\left(1+ \frac{r}{l}\right)+
\frac{B}{r^2},\label{nu'a}
\end{align}
with $B$ as an integration constant.

The integration of Eq.(\ref{nu'a}) yields
\begin{align}
\nu_a=-\frac{A}{r}e^{-r/l}- \frac{B}{r},\label{nua}
\end{align}

using the boundary condition to be zero in infinity. The specific
form of the integration constants $A$ and $B$ depends on the density
distribution of the galaxy, i.e., on the inner solution of the field
equations, according to the continuity
conditions at the surface $r=R_0$.\\\\

\item The fields for $r\leq R_0$ and the boundary and continuity conditions:\\
In the case $\hat{{\cal q}}=1$ the energy-momentum law (\ref{ideal})
is the same as in General Relativity. Octavio Obregón calculated in
1971 the general relativistic barometric formulae of matter in the
case of hydrostatic equilibrium and found for the coupling of
pressure and density of polytropic systems with the equation of
state $p=\alpha \varrho^\gamma$ (polytropic amplitude $\alpha$), the
density distribution \cite{Dehnen72}
\begin{align}
\varrho= \left(\frac{1}{\alpha}\right)^{\frac{1}{\gamma-
1}}\left[\left(\frac{\zeta}{\zeta_s}\right)^{\frac{\gamma-
1}{\gamma}}- 1\right]^{\frac{1}{\gamma- 1}}
\end{align}
with $\zeta=e^{\nu/2}$ as the absolute value of the time-like
Killing vector and $\zeta_s=e^{\nu_s/2}$ as its value at the surface
of the distribution of matter. For $|\nu|\ll 1$, the density
distribution reads
\begin{align}
\varrho= \left(\frac{1}{\alpha}\right)^{\frac{1}{\gamma-1}}
\left[\frac{1}{2}\frac{\gamma-1}{\gamma}(\nu_s-
\nu)\right]^{\frac{1}{\gamma-1}}.\label{varrho-Emden}
\end{align}
For $\gamma=2$, the differential equations
(\ref{EinlinA})-(\ref{TEinA}) become linear in $\nu$ concerning
$\varrho$, in accordance with the linearized Einstein theory. Then,
the polytropic equation becomes
\begin{align}
p=\alpha \varrho^2, \quad \varrho=\frac{1}{4\alpha} (\nu_s-
\nu).\label{eqstate}
\end{align}
Therefore, $p$ can then be neglected in Eqs.(\ref{EinlinA}),
(\ref{EinlinD}) and (\ref{TEinA}) in the linear approximation. The
pressure parameter $a$ and the polytropic amplitude $\alpha$ can be
fitted differently from galaxy to galaxy as well as the central mass
$M_S$ and the mass $M$ of the sphere or its radius $R_0$. The
fundamental length $l$, however, should have the same value for all
galaxies. This value may further be constrained from an analysis
within the Friedmann-Lemaître cosmology with Robertson-Walker
metric, too.\\

Of course, instead of with the polytropic equation of state one
might start also with the empirical mass distribution according to
the surface brightness of the galaxies. However, then the rotation
of the galaxies would have to be taken into account in order to
avoid unphysical equations of state, and the mass-to-light ratio
must be investigated in our scalar-tensor theory. This will be
performed in a further paper.\\

The exterior field components ($r>R_0$) are given through
Eqs.(\ref{xi-exterior}) and (\ref{nua}). For the inner field ($r\leq
R_0$), there is, according to (\ref{EinlinA}), (\ref{EinlinD}) and
(\ref{eqstate}):
\begin{alignat}{1}
\nu_i=& \left[\nu_s-
\frac{1}{r}\left(\tilde{A}\sin\left(k\frac{r}{l}\right)+
\tilde{B}\cos\left(k\frac{r}{l}\right)+ \tilde{C}\sinh\left(\kappa
\frac{r}{l}\right)+ \tilde{D}\cosh\left(\kappa \frac{r}{l}\right)\right)\right],\label{nui-allg}\\
\xi_i=& \frac{1}{r}\left[\tilde{A}\left(\delta-
k^2\right)\sin\left(k\frac{r}{l}\right)+ \tilde{B}\left(\delta-
k^2\right)\cos\left(k\frac{r}{l}\right)+ \right.\nn\\
&\left.+
\tilde{D}\left(\delta+\kappa^2\right)\cosh\left(\kappa\frac{r}{l}\right)+
\tilde{C}\left(\delta+ \kappa^2\right)
\sinh\left(\kappa\frac{r}{l}\right)\right]\label{xii-allg}
\end{alignat}
with the definitions
\begin{equation}
\left.
\begin{array}{c}
k^2= \frac{1}{2}\left[\delta- 1+ \sqrt{\delta^2+ \delta+ 1}\right]\\
\kappa^2= \frac{1}{2}\left[1- \delta+ \sqrt{\delta^2+ \delta+
1}\right]
\end{array}
 \right. \label{k-kappa}
\end{equation}
and
\begin{align}
\frac{\delta}{l^2}=\frac{8\pi G}{3\alpha}.\label{Delta-alpha}
\end{align}
$\tilde{A}$, $\tilde{B}$, $\tilde{C}$ and $\tilde{D}$ are
integration constants, essentially defined through the behaviour at
$r=0$ and the continuity conditions at the surface $r=R_0$. Because
in Eq.(\ref{TEinA}) the effective gravitational coupling parameter
$G(\xi)=G/(1+\xi)$ runs to zero for $r\rightarrow 0$ in the case of
a central singular mass $M_S$ (c.f. (\ref{xi-bnu})), this singular
mass has no direct influence on the behaviour of $\nu$. It is only
active on $\xi$, according to Eqs.(\ref{Diracid}) or (\ref{EinlinD})
(see Eqs.(\ref{xi-bnu}) and (\ref{nusB})), and herewith only
indirectly on $\nu$ with respect to (\ref{TEinA}). However, the
singular behaviour of $\xi$ for $r\rightarrow 0$ (see
Eq.(\ref{xi-bnu})) implies according to Eq.(\ref{TEinA}) no singular
behaviour of $\nu$ for $r\rightarrow 0$. Therefore, $\nu$ must be
regular for $r\rightarrow 0$. So, it follows from (\ref{nui-allg})
immediately
\begin{align}
\tilde{D}=-\tilde{B}.
\end{align}

The continuity conditions for $\nu$ and $\xi$ and their derivatives
with respect to $r$ at $r=R_0$ lead to
\begin{alignat}{1}
\tilde{A}& \sin\left(k\frac{R_0}{l}\right)+
\tilde{B}\left[\cos\left(k\frac{R_0}{l}\right)-
\cosh\left(\kappa\frac{R_0}{l}\right)\right]+
\tilde{C}\sinh\left(\kappa\frac{R_0}{l}\right)=0,\label{1-Stetig}\\
&\left\{\tilde{A}k \cos\left(k\frac{R_0}{l}\right)- \tilde{B}\left[k
\sin\left(k\frac{R_0}{l}\right)+ \kappa
\sinh\left(\kappa\frac{R_0}{l}\right)\right]+ \tilde{C}\kappa
\cosh\left(\kappa\frac{R_0}{l}\right)\right\}=l\nu_s- Ae^{-R_0/l},\label{2-Stetig}\\
&\left\{\tilde{A}k^2 \sin\left(k\frac{R_0}{l}\right)+
\tilde{B}\left[k^2 \cos\left(k\frac{R_0}{l}\right)+
\kappa^2\cosh\left(\kappa\frac{R_0}{l}\right)\right]-
\tilde{C}\kappa^2 \sinh\left(\kappa \frac{R_0}{l}\right)\right\}=
-Ae^{-R_0/l},\label{3-Stetig}\\
&\left\{\tilde{A}k^3\cos\left(k\frac{R_0}{l}\right)-
\tilde{B}\left[k^3 \sin\left(k\frac{R_0}{l}\right)- \kappa^3
\sinh\left(\kappa \frac{R_0}{l}\right)\right]- \tilde{C}
\kappa^3 \cosh\left(\kappa \frac{R_0}{l}\right)\right\}=\nn\\
&\quad= \delta l \nu_s+ A(1- \delta)e^{-R_0/l},\label{4-Stetig}
\end{alignat}
where $\nu_s$ is given by (see Eq.(\ref{nua})):
\begin{align}
\nu_s= -\frac{A}{R_0}e^{-R_0/l}- \frac{B}{R_0}.\label{nus-B}
\end{align}

From Eq.(\ref{1-Stetig}) it follows immediately
\begin{align}
\tilde{A}=\tilde{B}\frac{\cosh\left(\kappa \frac{R_0}{l}\right)-
\cos\left(k\frac{R_0}{l}\right)}{\sin\left(k\frac{R_0}{l}\right)}-
\tilde{C}\frac{\sinh\left(\kappa\frac{R_0}{l}\right)}{\sin
\left(k\frac{R_0}{l}\right)}.\label{A-B-C}
\end{align}
On the other hand, the $\xi_i$ field becomes for $r\rightarrow 0$,
according to Eq.(\ref{xii-allg}):
\begin{align}
\xi_i\rightarrow -\frac{\tilde{B}}{r}\left(k^2+
\kappa^2\right).\label{xi-bnu}
\end{align}
The integration constant $\tilde{B}$ can be coupled to the central
mass, using the equation of state (\ref{extreme}) for the centre of
the polytropic sphere; equation (\ref{EinlinD}) can then be
rewritten:
\begin{align}
\Delta \xi- \frac{1}{l^2}\xi=- \frac{8\pi
G}{3}[M_S\delta(\vec{x})(1- 3a)+ \varrho]\label{centrep}
\end{align}
with the pressure parameter $a$ and central mass $M_S$. The
connection between $\tilde{B}$ and the singular mass $M_S$ is found
through integration of Eq.(\ref{centrep}) for the sphere of radius
$r$ and then taking the limit $r\rightarrow 0$. In
\begin{align}
\lim_{r\rightarrow 0} \int_V \left(\Delta \xi- \frac{1}{l^2}\xi+
\frac{8\pi G}{3}\varrho\right)dV\label{limint}
\end{align}
the term $\lim_{r\rightarrow 0} \int_V (\frac{1}{l^2}\xi- \frac{8\pi
G}{3}\varrho)dV$ does not contribute, since the integral goes to
zero as $r^2$. The volume integral of $\Delta \xi$ can be
transformed with the help of the Gauss theorem to a surface
integral. Hence, with the solution (\ref{xi-bnu}), we get from
Eq.(\ref{centrep}):
\begin{alignat}{1}
\lim_{r\rightarrow 0} \int_V \Delta \xi dV &= \lim_{r\rightarrow 0}
\oint_F \nabla \xi df=\nn\\
&= 4\pi \tilde{B} \left(k^2+ \kappa^2\right)=-\frac{8\pi G}{3}M_s(1-
3a).
\end{alignat}
Thus, we have
\begin{align}
\tilde{B}= -\frac{2}{3} G M_S \frac{1- 3a}{k^2+
\kappa^2}.\label{nusB}
\end{align}

Furthermore, an expression for the mass of the polytropic sphere is
necessary. From Eqs.(\ref{eqstate}) and (\ref{nui-allg}), it
follows:
\begin{alignat}{1}
M= \frac{\pi}{\alpha} \int_0^R &(\nu_s- \nu)r^2dr=
\frac{\pi}{\alpha}
\left[\tilde{A}\left(\left(\frac{l}{k}\right)^2\sin\left(k\frac{R_0}{l}\right)-
R_0\frac{l}{k}\cos\left(k\frac{R_0}{l}\right)\right)+\right.\nn\\
&+\tilde{B}\left(\left(\frac{l}{k}\right)^2\cos\left(k\frac{R_0}{l}\right)+
R_0\frac{l}{k}\sin\left(k\frac{R_0}{l}\right)-
\left(\frac{l}{k}\right)^2\right)+
\tilde{C}\left(R_0\frac{l}{\kappa}\cosh\left(\kappa
\frac{R_0}{l}\right)\right.-\nn\\
&- \left.\left.\left(\frac{l}{\kappa}\right)^2\sinh\left(\kappa
\frac{R_0}{l}\right)\right)-\tilde{B}\left(\frac{R_0\kappa}{l}-
\left(\frac{l}{\kappa}\right)^2\cosh\left(\kappa
\frac{R_0}{l}\right)+
\left(\frac{l}{\kappa}\right)^2\right)\right].\label{Masse}
\end{alignat}

Herewith, all integration constants are determined. They can be put
into Eq.(\ref{nui-allg}) and (\ref{xii-allg}) and in the vacuum
solutions (\ref{xi-exterior}) and (\ref{nua}).\\
The four equations (\ref{1-Stetig})-(\ref{4-Stetig}) and the two
equations (\ref{nusB}) and (\ref{Masse}) determine the parameters
$\tilde{A}$, $\tilde{B}$, $\tilde{C}$, $A$, $B$ and $R_0$ in
dependence of the variable masses $M$ and $M_S$; the quantities $a$,
$\alpha$ and $l$ are to be considered as fundamental natural
constants, the values of which are unknown of course, and which can
be defined in a suitable way. Obviously, there exists in general a
mass-radius relation $R_0(M,M_S)$.\\
The resolution of Eqs.(\ref{2-Stetig})-(\ref{A-B-C}) gives the
following results: $\tilde{B}$ is determined immediately, according
to Eq.(\ref{nusB}) by the value of $M_S$. For the other coefficients
it follows then successively, for $A$:
\begin{alignat}{1}
\frac{Ae^{-R_0/l}}{k^2+ \kappa^2}&\left[k^2+ \kappa^2+ \left(k^3-
\delta k\right)\cot\left(k\frac{R_0}{l}\right)+ \left(\kappa^3+
\delta
\kappa\right) \coth\left(\kappa \frac{R_0}{l}\right)\right]\nn\\
&=\tilde{B}\left[\left(\kappa^3+ \delta \kappa\right)
\sinh\left(\kappa \frac{R_0}{l}\right) \left(1- \coth^2\left(\kappa
\frac{R_0}{l}\right)\right)-\right.\nn\\
&\quad \left.- \left(k^3- \delta k\right) \sin\left(k
\frac{R_0}{l}\right) \left(1+ \cot^2\left(k
\frac{R_0}{l}\right)\right)\right].\label{A-Gleichung}
\end{alignat}
Then, for $B$:
\begin{alignat}{1}
\frac{Bl}{R_0}=& A\left[\frac{e^{-R_0/l}}{k^2+ \kappa^2}\left(k
\cot\left(k\frac{R_0}{l}\right)- \kappa \coth\left(\kappa
\frac{R_0}{l}\right)\right)- \left(1+
\frac{l}{R_0}\right)e^{-R_0/l}\right]+\nn\\
&+ \tilde{B}\left[k \sin\left(k\frac{R_0}{l}\right)+ \kappa
\sinh\left(\kappa \frac{R_0}{l}\right)+ k\left(
\cos\left(k\frac{R_0}{l}\right)-
\cosh\left(\kappa\frac{R_0}{l}\right)\right)\cot\left(k\frac{R_0}{l}\right)+ \right.\nn\\
&\left.+ \cosh\left(\kappa\frac{R_0}{l}\right) \left(k
\cot\left(k\frac{R_0}{l}\right) -\kappa \coth\left(\kappa
\frac{R_0}{l}\right)\right)\right].\label{B-Gleichung}
\end{alignat}
Finally:
\begin{alignat}{1}
\tilde{A}=& -\frac{Ae^{-R_0/l}}{k^2+ \kappa^2}
\csc\left(k\frac{R_0}{l}\right)-
\tilde{B}\cot\left(k\frac{R_0}{l}\right),\label{AA-Gleichung}\\
\tilde{C}=& \frac{Ae^{-R_0/l}}{k^2+ \kappa^2}\text{csch}\left(\kappa
\frac{R_0}{l}\right)+ \tilde{B}\coth\left(\kappa
\frac{R_0}{l}\right).\label{CC-Gleichung}
\end{alignat}
The value of $R_0$ is finally determined by Eq.(\ref{Masse}) through
the mass $M$ of the polytropic sphere, so that a mass-radius
relation exists. However, this procedure is correct so long as
$B\neq 0$, i.e. $M_S\neq 0$ (see Eq.(\ref{nusB})).\\
A totally different situation exists in the case of $\tilde{B}=0$,
i.e. $M_S=0$. Then it follows from Eq.(\ref{A-Gleichung}):
\begin{align}
k^2+ \kappa^2+ (k^3- \delta k)\cot\left(k\frac{R_0}{l}\right)+
(\kappa^3+ \delta \kappa)\coth\left(\kappa\frac{R_0}{l}\right)=0.
\end{align}
By this equation the radius $R_0$ is \emph{determined independently
from the mass $M$} only by the natural parameters $\alpha$, $l$ and
$G$.\footnote{The same situation is realized in Newton's theory in
case of the polytropic index 2.} Furthermore, one finds from
Eqs.(\ref{A-Gleichung})-(\ref{CC-Gleichung}):
\begin{alignat}{1}
\frac{Bl}{R_0}=& A\left[\frac{e^{-R_0/l}}{k^2+ \kappa^2}\left(k
\cot\left(k\frac{R_0}{l}\right)- \kappa \coth\left(\kappa
\frac{R_0}{l}\right)\right)- \left(1+
\frac{l}{R_0}\right)e^{-R_0/l}\right],\\
\tilde{A}=& -\frac{Ae^{-R_0/l}}{k^2+
\kappa^2}\csc\left(k\frac{R_0}{l}\right),\\
\tilde{C}=& \frac{Ae^{-R_0/l}}{k^2+ \kappa^2}\text{csch}\left(\kappa
\frac{R_0}{l}\right).
\end{alignat}
The value of $A$ is then finally determined by the mass $M$,
according to Eq.(\ref{Masse})
\end{itemize}

\section{Tangential velocity and Rotation curves}

For the tangential velocity $v_{tan}=v^\varphi$ for central
symmetry, it is valid \cite{Matos00}
\begin{align}
v^2_{tan}\equiv v^2=\frac{r\nu'}{2}.\label{v2}
\end{align}

Accordingly, the square velocity (\ref{v2}) is:
\begin{alignat}{1}
&\text{Inner region ($r\leq R_0$): }\nn\\
v_i^2&= \frac{1}{2r}\left[\left(\tilde{B}-
\tilde{A}\frac{k}{l}r\right)\cos\left(\frac{kr}{l}\right)-
\left(\tilde{B}+ \tilde{C}\frac{\kappa}{l} r\right)
\cosh\left(\frac{\kappa r}{l}\right)+
\right.\nn\\
&\left.+ \left(\tilde{A}+
\tilde{B}\frac{k}{l}r\right)\sin\left(\frac{kr}{l}\right)+
\left(\tilde{C}+ \tilde{B}\frac{\kappa}{l}
r\right)\sinh\left(\frac{\kappa
r}{l}\right)\right].\\
&\text{Outer region ($r>R_0$): }\nn\\
v_a^2&= \frac{e^{-r/l}}{2r}\left[A\left(1+ \frac{r}{l}\right)+
Be^{r/l}\right].
\end{alignat}
The integration constants $A$, $B$, $\tilde{A}$, $\tilde{B}$ and
$\tilde{C}$ are given above (c.f. Eqs.(\ref{Masse}) and
(\ref{A-Gleichung})-(\ref{CC-Gleichung})). For them, the
gravitational coupling constant is $G=\frac{3}{4}G_N$ with $G_N$ as
the Newtonian gravitational constant, because according to this
theory, the determination of the gravitational constant in the
laboratory \cite{Faessler05} is performed under the condition $l\gg$
diameter of the torsion balance.\\\\

An estimation of the polytropic amplitude with the use of the
caloric equation of state and the virial theorem leads to an
approximate value:
\begin{align}
\alpha \approx \frac{G_N R_0^2}{c^4},\quad \frac{\delta}{c^4}\approx
\left(\frac{l}{R_0}\right)^2.
\end{align}
With the Milky Way radius $R_1=2.5$kpc, it follows $\alpha  \approx
2.5\cdot 10^{29}\frac{s^2 m}{kg}/c^4$, which is of the right order
of magnitude according to the figures below.\\
The best fits for rotation curves of some galaxies are, in case of a
non-vanishing central mass $M_S$ are found under figures 1 through 6.\\
In case of a vanishing central mass, i.e. $M_S=0$, the rotation
curves for the same galaxies are found under figures 7 through 12.
\newpage

 \begin{figure}[h!] \centering
\includegraphics*[width=16.5cm]{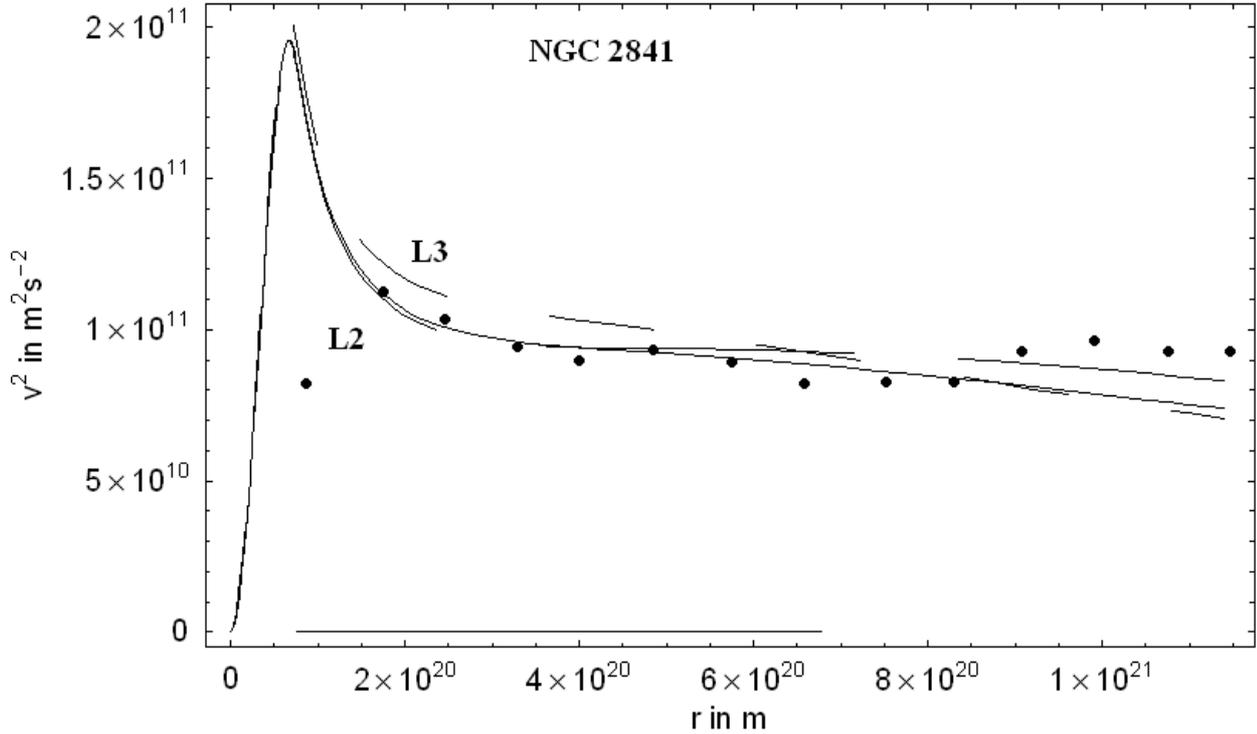}
\caption{Rotation curves with a length scale $l=6R_1$ (L2), $l=5R_1$
(L1) and $l=4R_1$ (L3) with experimental points for the galaxy NGC
2841 with $R_0=2.45$kpc \cite{Sanders86}. The galaxy mass is given
as $M=8.5\cdot 10^{10}M_{sun}$, with a central mass $M_S/M=180$, 140
and 120, respectively, whereas the polytropic amplitude varies
slowly according to $\alpha \cdot 10^{-30}\approx 0.27$, 0.27 and
$0.28 \frac{s^2 m}{kg}/c^4$, respectively. $R_1$ is the radius of
the Milky Way.}
 \end{figure}
 \begin{figure}[h!] \centering
\includegraphics*[width=16.7cm]{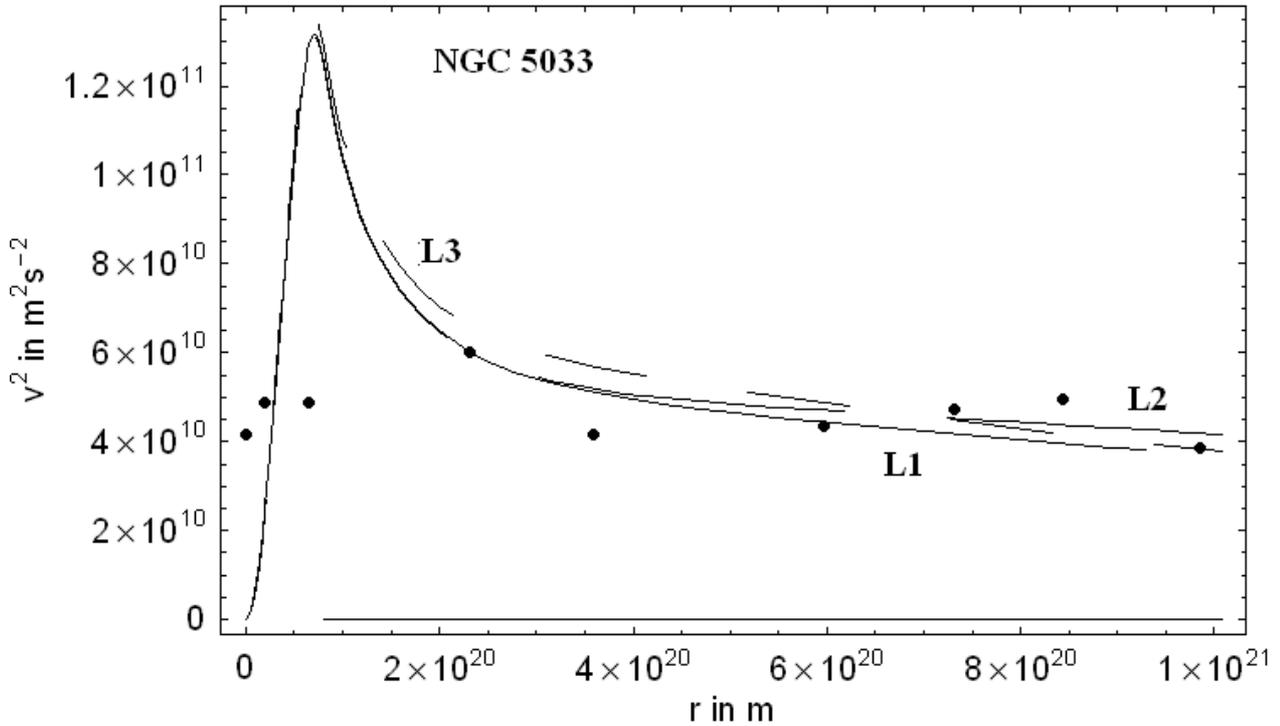}
\caption{Rotation curves with a length scale as in Figure 1 and
experimental points for the galaxy NGC 5033 \cite{Sanders90}. There
is $M=7\cdot 10^{10}M_{sun}$. The galaxy radius is $R_0=2.6$kpc. The
central mass is chosen as $M_S/M=80$, 60 and 55, respectively,
whereas the polytropic amplitude varies slowly according to $\alpha
\cdot 10^{-30}\approx 0.29$, 0.29 and $0.30 \frac{s^2 m}{kg}/c^4$,
respectively.}
\end{figure}
 \begin{figure}[h!] \centering
\includegraphics*[width=16.8cm]{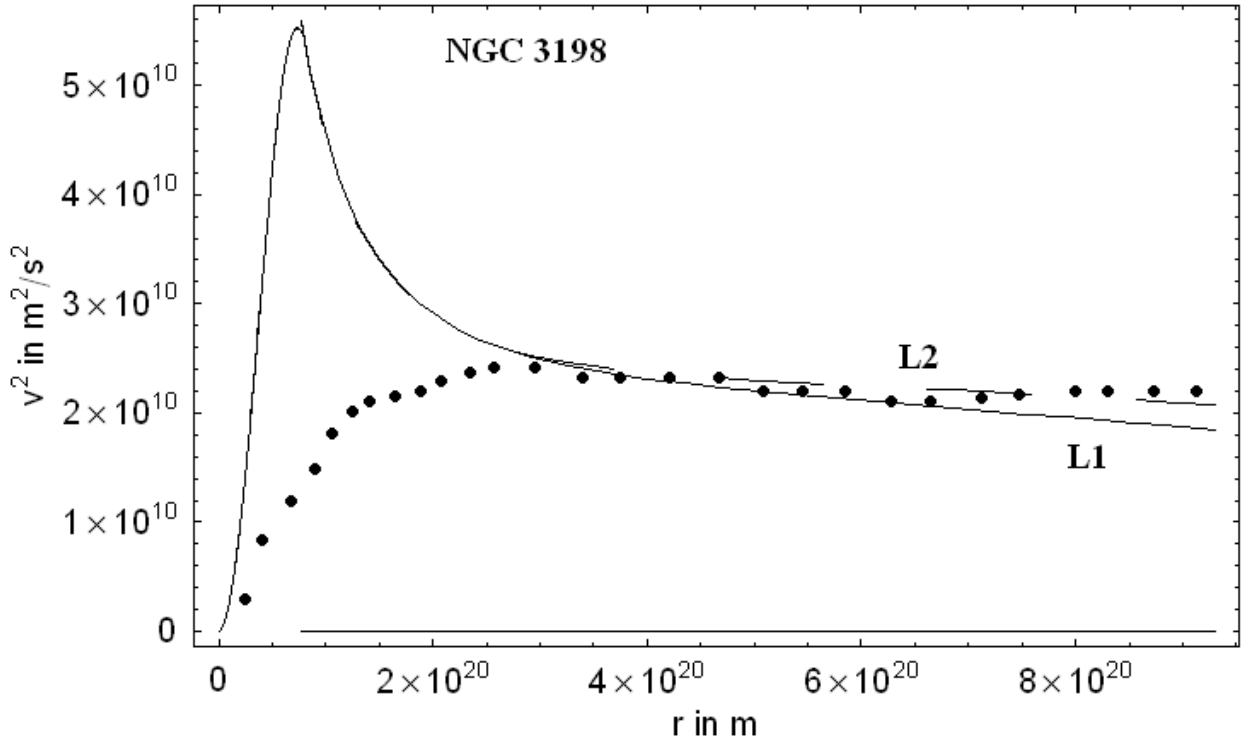}
\caption{Rotation curves with a length scale as in Figure 1 and
experimental points for the galaxy NGC 3198 with $R_0=2.68$kpc
\cite{Sanders86}. The galaxy mass is taken as $M=3 \cdot
10^{10}M_{sun}$ and the central mass $M_S/M=40$, and 30,
respectively, whereas the polytropic amplitude is $\alpha \cdot
10^{-30}\approx 0.32 \frac{s^2 m}{kg}/c^4$.}
\end{figure}
 \begin{figure}[h!] \centering
\includegraphics*[width=16.5cm]{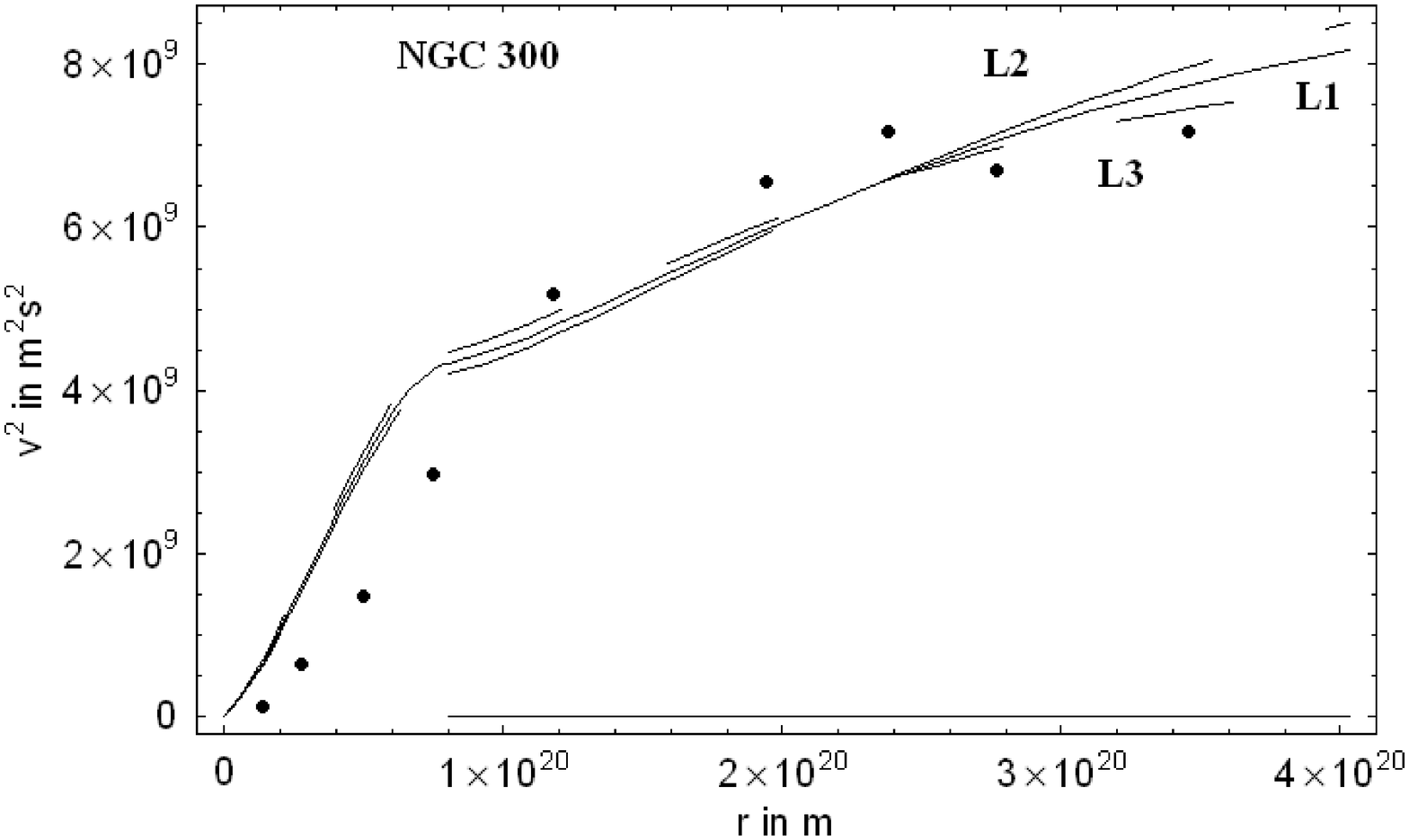}
\caption{Rotation curves with a length scale as in Figure 1 and
experimental points for the galaxy NGC 300 with $R_0=2.6$kpc
\cite{Sanders86}. There is $M=0.1 \cdot 10^{10}M_{sun}$ and
$M_S/M=23$, 17 and 12, respectively, whereas the polytropic
amplitude varies slowly according to $\alpha \cdot 10^{-30}\approx
0.63$, 0.64 and $0.67 \frac{s^2 m}{kg}/c^4$, respectively.}
 \end{figure}
\begin{figure}[h!] \centering
\includegraphics*[width=16.6cm]{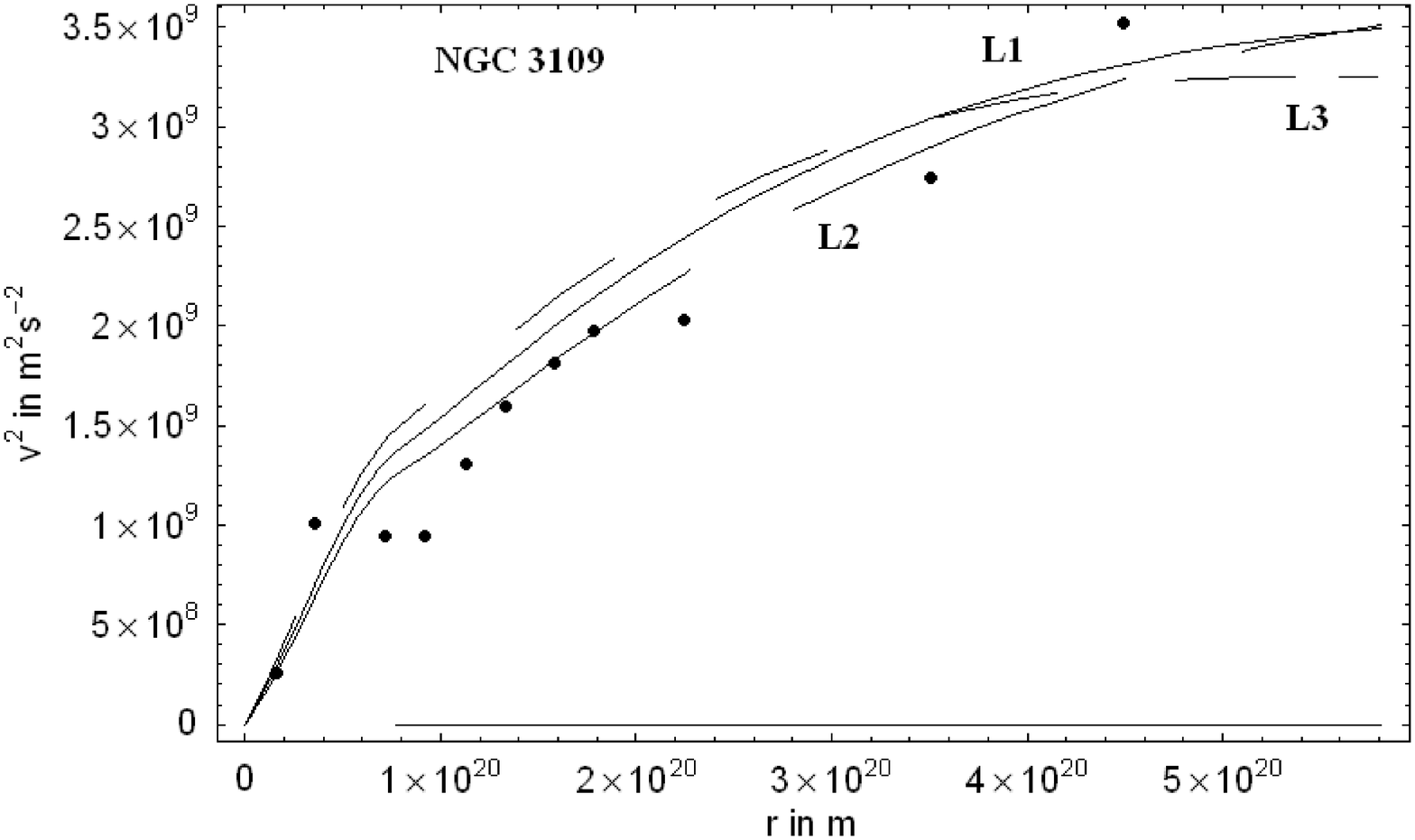}
\caption{Rotation curves with a length scale as in Figure 1 and
experimental points for the galaxy NGC 3109 with $R_0=2.5$kpc
\cite{Sanders86}. There is $M=0.02 \cdot 10^{10}M_{sun}$ and
$M_S/M=8.5$, 6.8 and 5, respectively, whereas the polytropic
amplitude varies slowly according to $\alpha \cdot 10^{-30}\approx
0.82$, 0.89 and $0.97 \frac{s^2 m}{kg}/c^4$, respectively.}
 \end{figure}
\begin{figure}[h!] \centering
\includegraphics*[width=16.7cm]{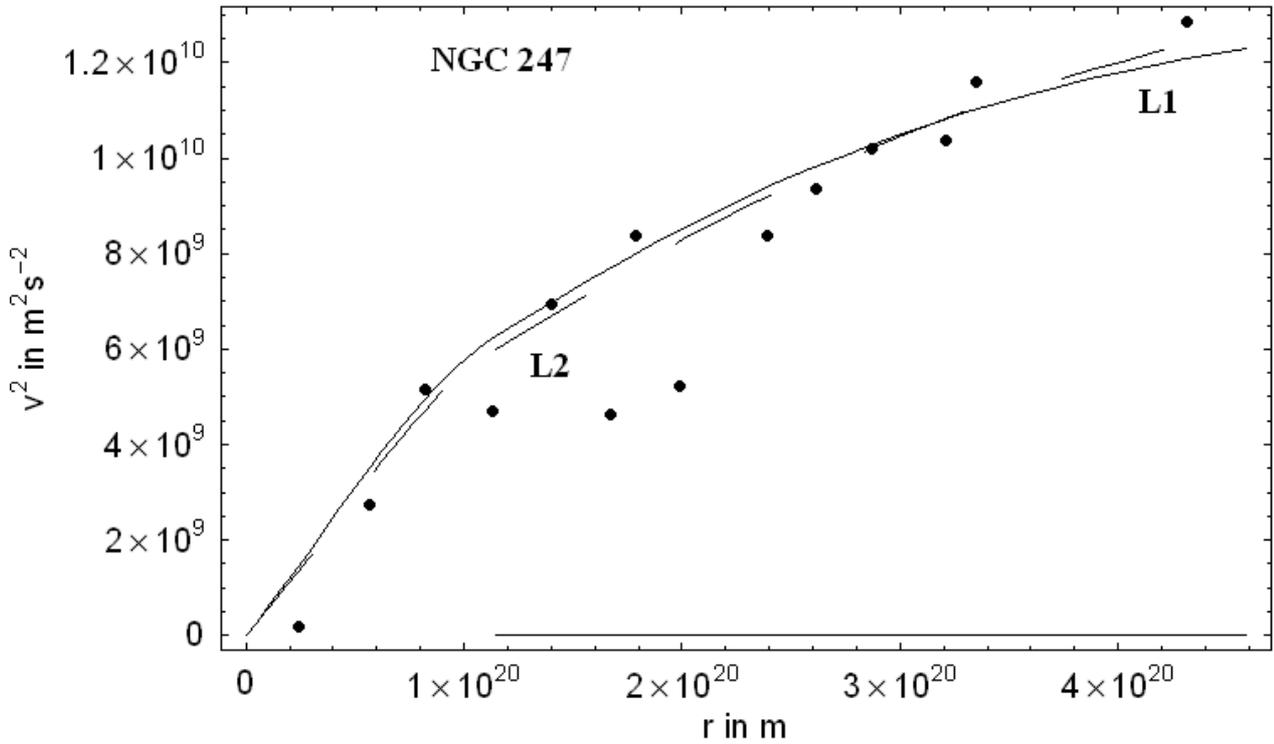}
\caption{Rotation curves with a length scale as in Figure 1 and
experimental points for the galaxy NGC 247 with $R_0=3.7$kpc
\cite{Sanders86}. There is $M=0.09 \cdot 10^{10}M_{sun}$ and
$M_S/M=33$ and 25, respectively, whereas the polytropic amplitude
varies slowly according to $\alpha\cdot 10^{-30}\approx 2.80$ and
$2.96 \frac{s^2 m}{kg}/c^4$, respectively, with a greater value as
in other figures because of $R_0>R_1$.}
 \end{figure}
\begin{figure}[h!] \centering
\includegraphics*[width=17.cm]{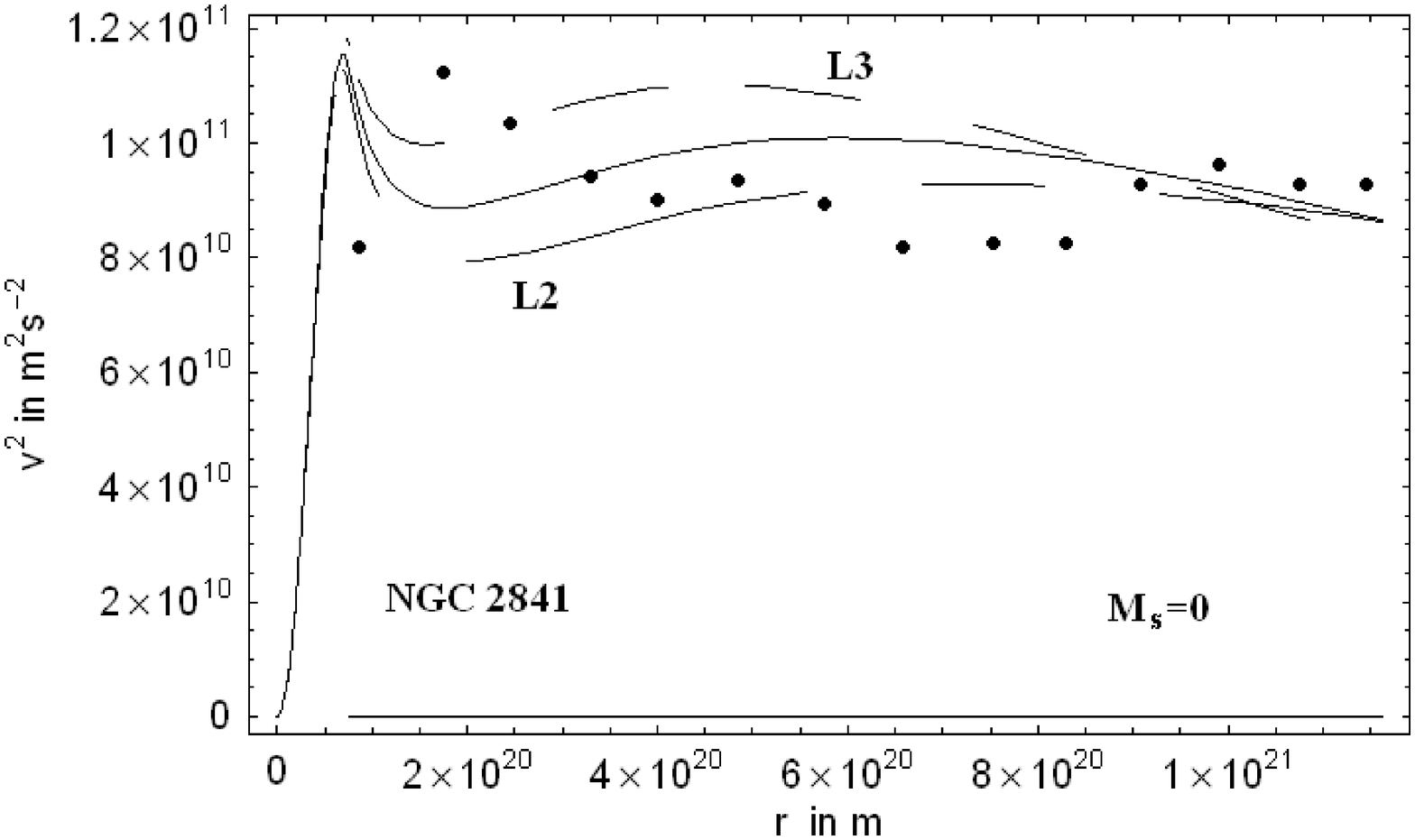}
\caption{Rotation curves with a length scale $l=6R_1$ (L2) and
$l=5R_1$ (L1) and $l=4R_1$ (L3) with experimental points for the
galaxy NGC 2841 with $R_0=2.45$kpc \cite{Sanders86}. The galaxy mass
is given as $M=5.5 \cdot 10^{10}M_{sun}$ and the central mass is
chosen as vanishing, whereas the polytropic amplitude varies slowly
according to $\alpha \cdot 10^{-30}\approx 0.27$, 0.27 and $0.28
\frac{s^2 m}{kg}/c^4$, respectively. $R_1$ is the radius of the
Milky Way.}
 \end{figure}
 \begin{figure}[h!] \centering
\includegraphics*[width=17.2cm]{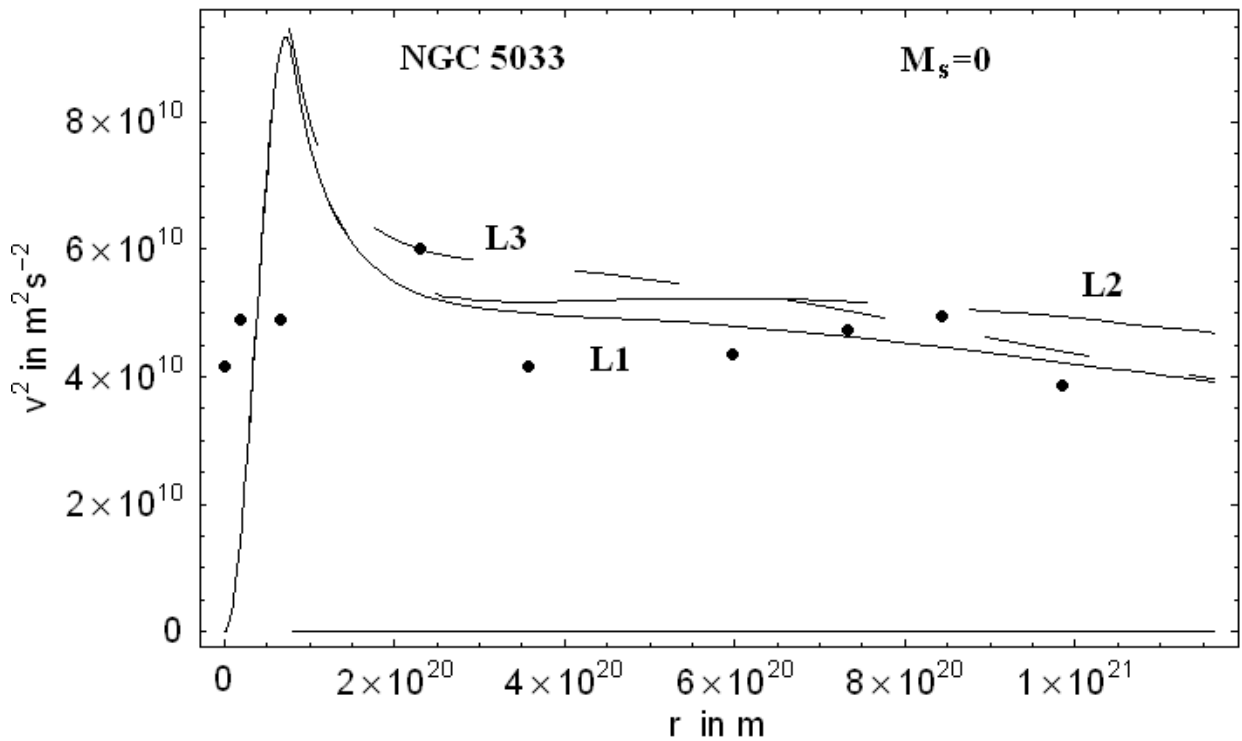}
\caption{Rotation curves with a length scale as in Figure 7 and
experimental points for the galaxy NGC 5033 \cite{Sanders90} and
radius $R_0=2.6$kpc. There is $M=5 \cdot 10^{10}M_{sun}$ with a
vanishing central mass, whereas the polytropic amplitude varies
slowly according to $\alpha \cdot 10^{-30}\approx 0.29$, 0.29 and
$0.30 \frac{s^2 m}{kg}/c^4$, respectively.}
\end{figure}
 \begin{figure}[h!] \centering
\includegraphics*[width=17.2cm]{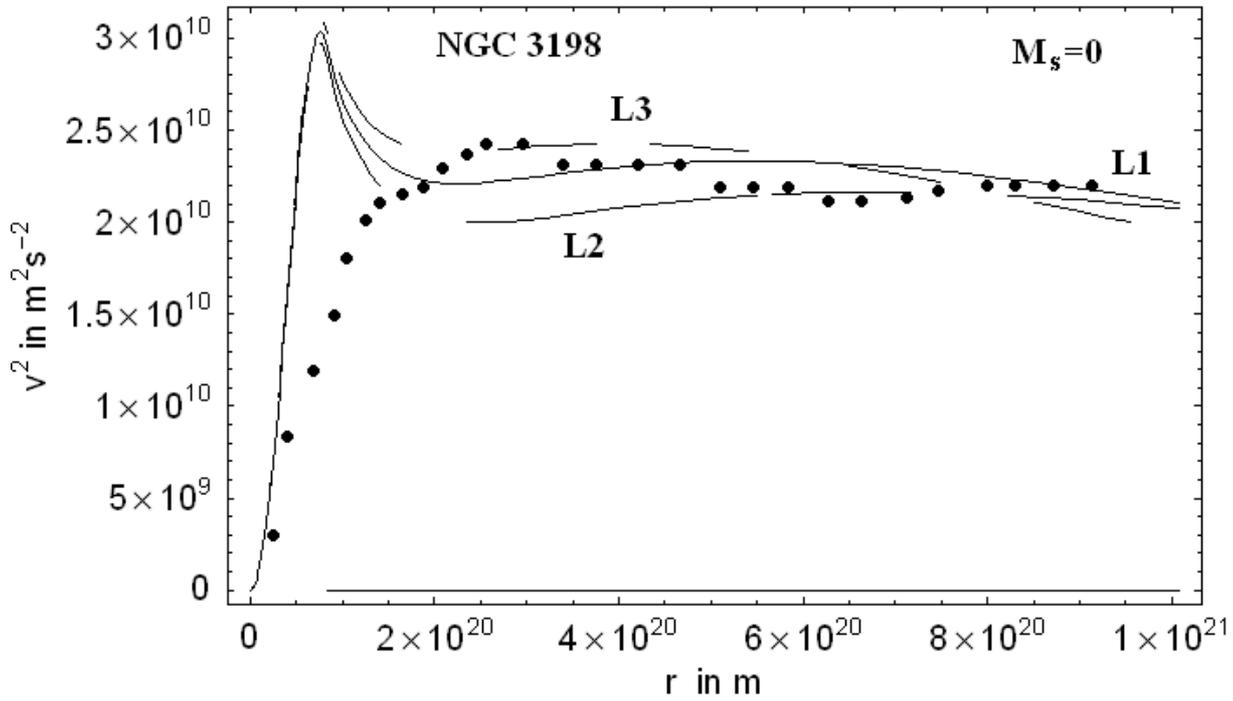}
\caption{Rotation curves with a length scale as in Figure 7 and
experimental points for the galaxy NGC 3198 with $R_0=2.68$kpc
\cite{Sanders86}. There is $M=3 \cdot 10^{10}M_{sun}$ and $M_S=0$,
whereas $\alpha \cdot 10^{-30}\approx 0.32 \frac{s^2 m}{kg}/c^4$.}
\end{figure}
 \begin{figure}[h!] \centering
\includegraphics*[width=17.2cm]{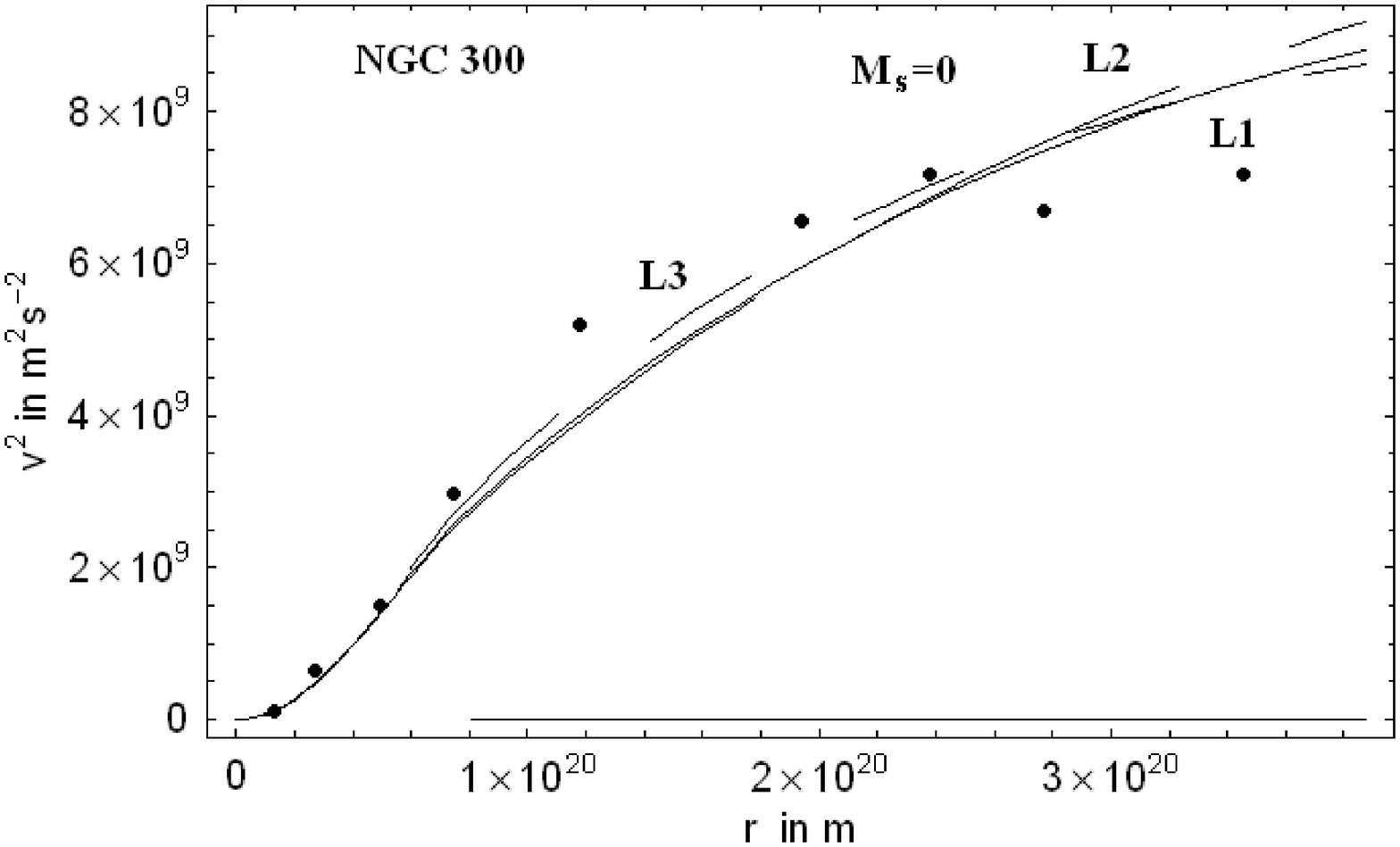}
\caption{Rotation curves with a length scale as in Figure 7 and
experimental points for the galaxy NGC 300 with $R_0=2.6$kpc
\cite{Sanders86}. There is $M=0.055 \cdot 10^{10}M_{sun}$ and
$M_S=0$, whereas the polytropic amplitude varies slowly according to
$\alpha \cdot 10^{-30}\approx 0.63$, 0.64 and $0.67 \frac{s^2
m}{kg}/c^4$, respectively.}
 \end{figure}
\begin{figure}[h!] \centering
\includegraphics*[width=16.7cm]{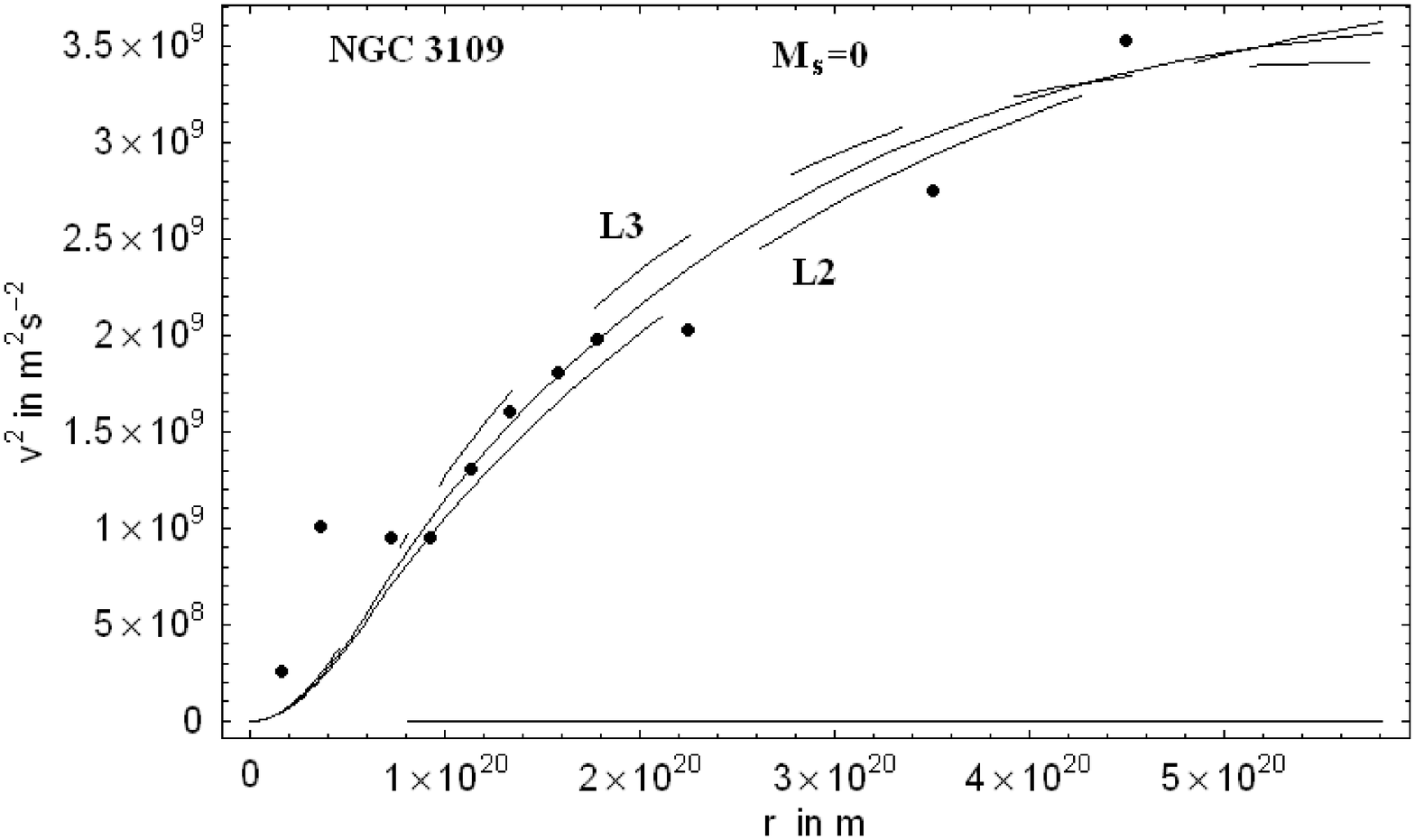}
\caption{Rotation curves with a length scale as in Figure 7 and
experimental points for the galaxy NGC 3109 with $R_0=2.5$kpc
\cite{Sanders86}. There is $M=0.01 \cdot 10^{10}M_{sun}$ and
$M_S=0$, whereas the polytropic amplitude varies slowly according to
$\alpha \cdot 10^{-30}\approx 0.83$, 0.89 and $0.97 \frac{s^2
m}{kg}/c^4$, respectively.}
 \end{figure}
\begin{figure}[h!] \centering
\includegraphics*[width=16.9cm]{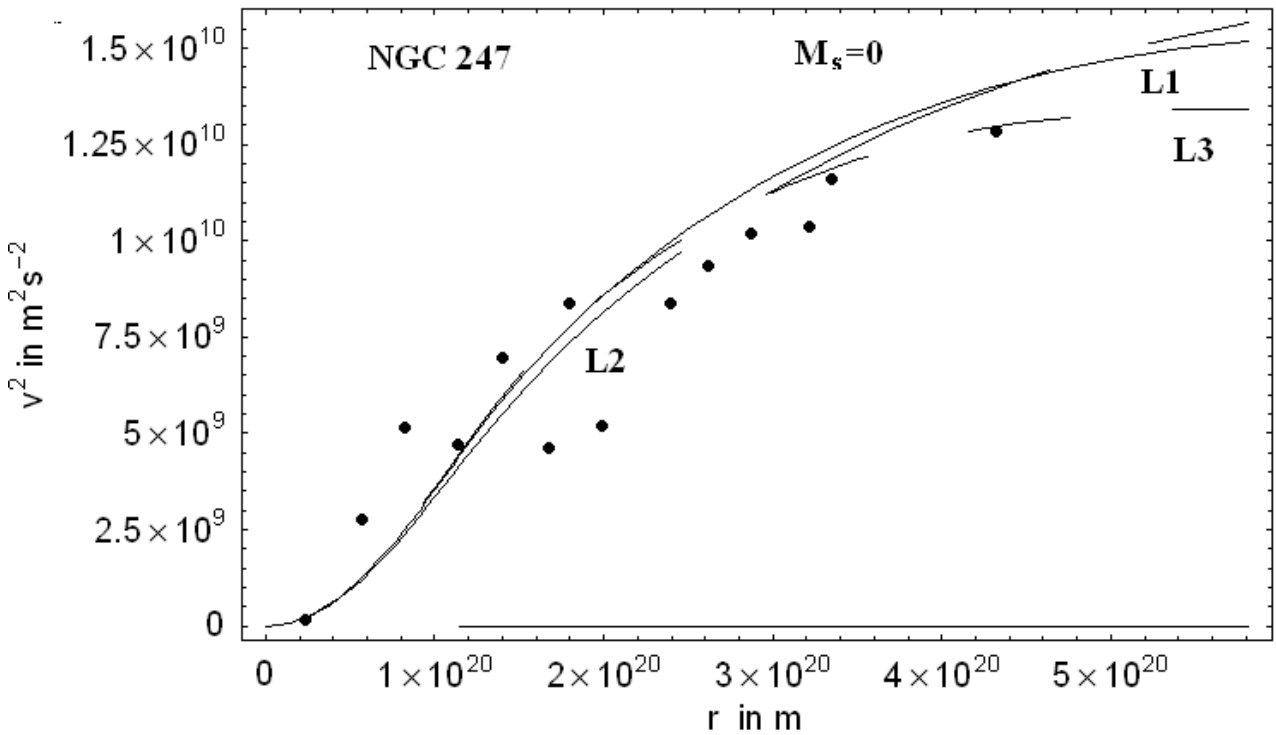}
\caption{Rotation curves with a length scale as in Figure 7 and
experimental points for the galaxy NGC 247 with $R_0=3.7$kpc
\cite{Sanders86}. There is $M=0.05 \cdot 10^{10}M_{sun}$ and
$M_S=0$, whereas the polytropic amplitude varies slowly according to
$\alpha\cdot 10^{-30}\approx 2.80$ and $2.96 \frac{s^2 m}{kg}/c^4$,
respectively (analog to Figure 6).}
 \end{figure}

\newpage

\section{Conclusions}
The comparison of the theoretical rotation curves with the rotation
curves for several galaxies indicates that the scalar-tensor theory
with the Higgs Mechanism is able to explain and contribute to the
explanation of their flatness, although in the case of high
tangential velocities (see Fig. 1-3 and 7-9) the theoretical curves
show a peak at $r=R_0$, which is observationally not always
verified. Furthermore, it follows that a polytropic density
distribution for galaxies is useful to achieve a satisfactory
agreement between theoretical and empirical data, postulating or not
postulating a central massive core for the galaxies. Nevertheless,
the dynamics seem better explained by a non-vanishing central mass,
which leads to the necessity of a higher galaxy mass. The central
mass lies in the order of several galaxy masses, but is in
conformity with this theory, since the mass interacts with the outer
region only very weakly. This is in contrast to Newtonian mechanics,
because in this scalar-tensor theory the effective gravitational
coupling parameter is drastically diminished in the centre.\\

\bibliography{apssamp}

\end{document}